\documentclass[conference]{IEEEtran}

\usepackage[utf8]{inputenc}
\usepackage[english]{babel}

\usepackage{booktabs}
\usepackage{multirow}

\usepackage{cite}
\usepackage{graphicx}
\usepackage{textcomp}
\usepackage{xcolor}

\def\BibTeX{{\rm B\kern-.05em{\sc i\kern-.025em b}\kern-.08em
    T\kern-.1667em\lower.7ex\hbox{E}\kern-.125emX}}

\usepackage{amsmath,amssymb,amsfonts}
\usepackage{bm}

\usepackage[cmintegrals]{newtxmath}

\newcommand\blfootnote[1]{%
  \begingroup
  \renewcommand\thefootnote{}\footnote{#1}%
  \addtocounter{footnote}{-1}%
  \endgroup
}

\begin{document}

\title{Beyond 5G Low-Power Wide-Area Networks:\\A LoRaWAN Suitability Study}

\author{\IEEEauthorblockN{
    Arliones~Hoeller\IEEEauthorrefmark{1}\IEEEauthorrefmark{2}\IEEEauthorrefmark{3},  
    Jean~Sant'Ana\IEEEauthorrefmark{1}, 
    Juho Markkula\IEEEauthorrefmark{1}, 
    Konstantin~Mikhaylov\IEEEauthorrefmark{1}, 
    Richard Souza\IEEEauthorrefmark{2}, 
    Hirley~Alves\IEEEauthorrefmark{1}
    }
    \IEEEauthorblockA{\IEEEauthorrefmark{1}6G Flagship, Centre for Wireless Communications, University of Oulu, Finland, Firstname.Lastname@oulu.fi\\}
    \IEEEauthorblockA{\IEEEauthorrefmark{2}Federal University of Santa Catarina, Florianópolis, Brazil, richard.demo@ufsc.br\\}
    \IEEEauthorblockA{\IEEEauthorrefmark{3}Federal Institute for Education, Science, and Technology of Santa Catarina, São José, Brazil, arliones.hoeller@ifsc.edu.br}\vspace{-.75cm}
}

\maketitle

\begin{abstract}
In this paper, we deliver a discussion regarding the role of Low-Power Wide-Area Networks (LPWAN) in the cellular Internet-of-Things (IoT) infrastructure to support massive Machine-Type Communications (mMTC) in next-generation wireless systems beyond 5G. We commence by presenting a performance analysis of current LPWAN systems, specifically LoRaWAN, in terms of coverage and throughput. The results obtained using analytic methods and network simulations are combined in the paper for getting a more comprehensive vision. Next, we identify possible performance bottlenecks, speculate on the characteristics of coming IoT applications, and seek to identify potential enhancements to the current technologies that may overcome the identified shortcomings.\blfootnote{Paper accepted for presentation at 6G Wireless Summit 2020.}\blfootnote{\textcopyright 2020 IEEE. Personal use of this material is permitted. Permission from IEEE must be obtained for all other uses, in any current or future media, including reprinting/republishing this material for advertising or promotional purposes, creating new collective works, for resale or redistribution to servers or lists, or reuse of any copyrighted component of this work in other works.}
\end{abstract}

\section{Introduction}

The 5G systems address three types of network services: Enhanced Mobile Broadband (eMBB), Ultra-Reliable Low-Latency Communications (URLLC), and massive Machine-Type Communications (mMTC).
The eMBB and URLLC services, although quite complex, focus on well known and extensively studied sets of applications.
The eMBB depends, basically, on human demand for high bandwidth, including, for instance, video streaming and video conference applications, and 5G boosts it by increasing spectrum efficiency to support more users at higher bit rates.
URLLC applications, although new in the context of cellular networks, relates to a well studied set of critical real-time applications, for which behavior is usually deterministic and formally specified, allowing 5G to deliver the service by thoroughly planning and managing the resource allocation.
The mMTC services, however, must cope with Ultra-Dense Networks (UDN) of devices with dynamic and sporadic traffic patterns~\cite{Kamel:TVT:2019}.
That poses challenges to delivering massive connectivity with acceptable reliability~\cite{Sharma:CST:2019} and promoting efficient resource utilization.

The Internet-of-Things (IoT) paradigm demands mMTC to serve \textit{massive numbers of users} with \textit{low-energy consumption} and \textit{reasonable reliability}.
The first two requirements are addressed by Low-Power Wide-Area Networks (LPWAN) like LoRaWAN, SigFox, and NB-IoT.
However, to achieve large scale connectivity and low energy consumption, LPWANs replace complicated channel control mechanisms by simpler MAC protocols, at the cost of reliability~\cite{Raza:CST:2017}.
Although LPWANs are in fast-paced adoption, reports on deployments with large numbers of stations are yet to come out. Therefore LPWAN performance and capacity models are still an open problem.
As a result, recent studies explored the capacity limits of the technologies and proposed techniques to enhance their performance.
In recent work, we studied, modeled, analyzed~\cite{Hoeller:Access:2019}, and simulated~\cite{Markkula:ICC:2019} the performance of the uplink of the LoRa technology, which is the physical layer of LoRaWAN~\cite{LoRaAlliance:2018}.
These works allowed us to understand some characteristics of LoRa networks and, through extrapolation, other LPWAN.

The LoRaWAN protocol stack emerges as a promising LPWAN solution for mMTC applications.
Its openness facilitates and encourages its adoption by both researchers and practitioners.
However, it may experience decreased reliability with the increase of the numbers of users~\cite{Mahmood:TII:2019}. To approach this challenge and improve the LPWAN performance, the scholars propose to exploit the diversity relying on independent realizations of the wireless channel.
For instance, time diversity is approached in the form of independent~\cite{Hoeller:Access:2018} or coded~\cite{Santana:TII:2020} message replications, or through spatial diversity~\cite{Hoeller:Access:2018}.
In order to improve mMTC performance in UDNs, authors have also shown that resource reuse and interference mitigation improve the efficiency of the Radio Access Network (RAN) significantly~\cite{Bartoli:IOTJ:2019}.
Also, it has been shown that Multi-Radio Massive Machine-Type Communication (MR-MMTC) systems improve reliability, coverage, latency, and throughput of MTC networks by adapting the communication front-end to cope with transient performance degradation~\cite{Mikhaylov:CM:2019}.
Besides that, Non-Orthogonal Multiple Access (NOMA) has been considered as a means to increase spectral efficiency~\cite{Mohammadkarimi:VTM:2018}.
Finally, the use of drone- and satellite-based gateways has also been considered to improve coverage~\cite{Sharma:Energies:2018,Cioni:NTW:2018}.

In this paper, we bring together our previous work on analysis~\cite{Hoeller:Access:2019} and simulation~\cite{Markkula:ICC:2019} of the performance of LoRaWAN systems to identify the major bottlenecks in terms of coverage and throughput.
We then consider possible enhancements of the technologies to cope with the expected increased demand from future IoT applications.

\section{LPWAN}\label{sec:baseline}

LPWAN is a category of wireless networks whose design challenges include long-range communication, low power consumption, low cost, scalability, and adaptable reliability~\cite{Raza:CST:2017}.
These networks serve IoT applications for which technologies like Bluetooth, WLAN, and Zigbee are inefficient.
The use of sub-GHz bands with narrowband (\textit{e.g.}, Sigfox) and spread spectrum modulations (\textit{e.g.}, LoRa, RPMA) increases the link budget and supports the utilization of a star topology.
This configuration reduces the volume of control messages, energy consumption, and latency~\cite{Raza:CST:2017}.
Also, the use of lightweight MAC protocols like ALOHA is common to reduce power consumption and hardware complexity and cost.

Nonetheless, the major downside is the network scalability. 
Some technologies promote signal diversity by its core protocols, as is the case, \textit{e.g.}, of the message replication in different frequencies by Sigfox, and the reception of uplink packets by multiple base stations in LoRaWAN and Sigfox.
Regarding reliability, some LPWANs can change their data rates to adapt to different reliability requirements (\textit{e.g.}, LoRa, NB-IoT).

Although several LPWAN technologies have emerged, LoRaWAN has been the most widely used LPWAN technology in the scientific community.
It happens because (i)~LoRaWAN is an open standard operating in the sub-GHz industrial, scientific and medical (ISM) radio bands, (ii)~LoRa end-devices and gateways are widely commercially available, and (iii)~it is possible to install and operate a LoRaWAN cell, virtually, anywhere.
Among the other most known technologies, Sigfox includes undisclosed proprietary technology and has a single service provider, and NB-IoT is just now making it to the general market.
Besides that, these LPWAN technologies feature similar characteristics, and, thus, some models and conclusions drawn for one technology can be extrapolated to the others.
Here, we consider LoRaWAN because of the easier access to the technology.
Besides that, our group operates a campus-wide LoRaWAN network at the University of Oulu, with which we acquired relevant experience with the technology.

\subsection{LoRaWAN}

LoRa is a proprietary PHY technology that uses Chirp Spreading Spectrum (CSS) modulation to spread narrowband signals into a wideband channel, generating processing gain. This gain depends on the used spreading factor (SF), which varies from 7 to 12. Higher SF results in lower data rate but greater processing gain. Also, SFs are quasi-orthogonal, enabling a gateway to decode multiple signals sent simultaneously with different SFs in the same frequency band. Typically, LoRa operates on sub-GHz ISM bands, which, depending on the region, inquire specific restrictions for the frequency bands and transmit power (\textit{e.g.}, 868MHz/14dBm for EU, 915MHz/21.7dBm for the US) and the maximum duty cycle (between 0.1\% and 10\%, usually 1\% in EU). Since LoRa is frequency modulatied, it demonstrates the capture effect, enabling a receiver to decode a colliding signal, given that this signal is sufficiently stronger than the interference.

LoRaWAN is an open protocol stack that uses LoRa as PHY~\cite{LoRaAlliance:2018} and is developed and maintained by the LoRa Alliance.
LoRaWAN features a star topology and employs three types of equipment.
The \textit{end-devices} send uplink messages to one or more \textit{gateways}, that forward them to a Network Server (\textit{netserver}) via an IP connection.
Devices can belong to three classes.
The basic class, Class A, uses pure-ALOHA channel access with a random selection of the frequency channel.
After each uplink, a device opens two receiving windows for downlink messages from the netserver.
In addition to this, class B devices synchronize with the netserver through periodic beacons and open receiving windows periodically.
Class C devices are usually mains-powered and keep the radio in receive whenever not transmitting.

\section{Performance of LoRaWAN\label{sec:models}}



In previous work, we addressed, analytically, the coverage probability of devices in LoRaWAN cells~\cite{Hoeller:Access:2019} and, by simulation, the throughput and packet delivery ratio (PDR) of such systems~\cite{Markkula:ICC:2019}.
In this section, we briefly introduce the models.

\subsection{Theoretical Models}


In LoRaWAN, devices usually employ the Adaptive Data Rate mechanism to set the SF of each device according to the channel condition measured at the gateway.
Since the channel condition depends on the communication distance, analytical models adopt a ring-based network topology, setting SF according to the device's distance from the gateway.

\begin{figure}[tb]
    \centering
    \includegraphics[width=.8\columnwidth]{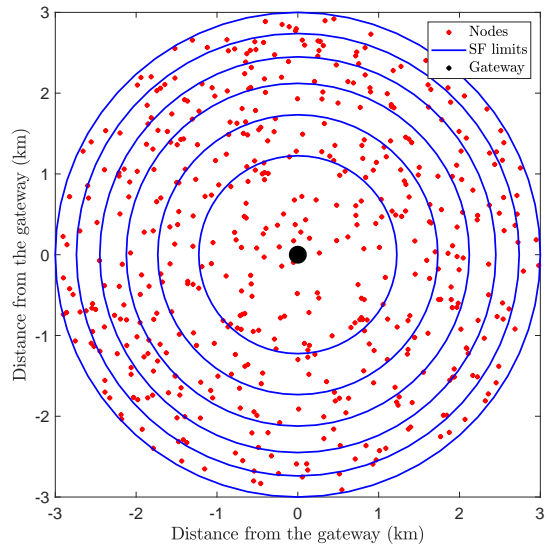}
    \caption{$\bar{N}=500$ nodes uniformly distributed in a circular area of radius $R=3000$~m around the gateway. SF allocated using equal-area rings.}
    \label{fig:nodes}
\end{figure}

Figure~\ref{fig:nodes} illustrates an \emph{exemplary} setup where SF increases with the distance from the gateway.
We consider $N_j$ nodes deployed uniformly inside the $j$-th circular ring of area $V_j=\pi(l_j^2-l_{j-1}^2)$, where $l_j$ and $l_{j-1}$ define the ring's outer and inner radii.
We assume that nodes always have a packet ready for transmission and that transmissions follow pure-ALOHA with probability $p$.
The model assumes a Poisson Point Process (PPP) for each $j$-th SF ring ($\Phi_j$) with intensity $\alpha_j=p\rho_j$, where $\rho_j=\overline{N}_j/{V_j}$ is the spatial density of nodes inside ring $j$.
The average number of nodes in the network is $\overline{N} = \sum \overline{N}_j$.
All nodes transmit with the same power $P_{tx}$.
A signal $r_1$ received from a typical node includes the transmitted signal ($s_1$), AWGN ($w$), and interference signals ($s_k$), attenuated by path loss ($g_k$) and Rayleigh fading ($h_k$), \textit{i.e.},
\begin{align}
    r_1 &= \sqrt{P_{tx}g_1}h_1*s_1 + \sum\limits_{j\in S} \sum\limits_{k\in\Phi_j} \sqrt{P_{tx}g_k}h_k*s_k + w,
\end{align}
where the path loss law is $g_k=\left(\frac{\lambda}{4\pi d_k}\right)^\eta$, with $\lambda$ as the wave-length, $d_k$ as the distance between the $k$-th node and the gateway, and $\eta>2$ is the path loss exponent.

The coverage probability is the product of the noise-dependent connection probability ($H_1$) and the interference-dependent capture probability ($Q_1$), \textit{i.e.},  $C_1=H_1Q_1$.
Following~\cite{Hoeller:Access:2019}, the connection probability is $H_1 = \mathbb{P}\left( \textup{SNR} > \gamma_i \right)$, and
\begin{align}
    H_1 = \mathbb{P}\left( \frac{P_{tx}g_1|h_1|^2}{\sigma_w^2} > \gamma_i \right) = \exp\left(-\frac{\gamma_i\sigma_w^2}{P_{tx}g_1}\right),\label{eqn:h1}
\end{align}
with $\sigma_w^2$ as the variance of the AWGN and $i$ as the SF ring of the typical node.
The SNR threshold $\gamma_i$ differs for each SF $i$.

In~\cite{Hoeller:Access:2019}, the capture probability considers the interference-limited communication probability considering both intra-SF and inter-SF interference sources.
The capture probability is analyzed separately for the SIR$_j$ from each SF ring $j$.
By using stochastic geometry~\cite{Haenggi:Book:2012} to model interference, \cite{Hoeller:Access:2019}~proposes that the capture probability for a given SF ring $j$ is $P_{SIR_j}=\mathbb{P}(\textup{SIR}_j>\delta_{ij})$, where $\delta_{ij}$ is the SIR threshold for decoding an SF $i$ signal with interference from SF $j$.
It results in
\begin{align}
    P_{SIR_j} = \exp\left\{ -\pi\alpha_j \left[ l_j^2\,_2F_1\left(1,\frac{2}{\eta};1+\frac{2}{\eta};-\frac{l_j^\eta}{d_1^\eta\delta_{ij}}\right) \right.\right.\nonumber\\
    \left.\left. - l_{j-1}^2\,_2F_1\left(1,\frac{2}{\eta};1+\frac{2}{\eta};-\frac{l_{j-1}^\eta}{d_1^\eta\delta_{ij}}\right) \right]  \right\}.
\end{align}

Finally, an outage takes place if the SIR for at least one interfering SF exceeds its respective threshold.
Conversely, the probability that a collision does not occur is $Q_1 = \prod_{j\in S} P_{\textup{SIR}_j}$, and since $P_{\textup{SIR}_j}$ is exponential,
\begin{align}
    Q_1 = \textup{exp} \left ( \sum_{j \in S} P_{SIR_j} \right ). \label{eqn:q1}
\end{align}

\subsection{LoRaWAN simulator}

In \cite{Markkula:ICC:2019}, the LoRaWAN simulation model for the Riverbed Modeler network simulator is delivered.
The simulation model is based on the Hata Rural path loss model~\cite{Molisch:Book:2011}, LoRaWAN transceivers that apply distinct bit error rate (BER) curves for different SF, the pure-ALOHA of the Class A end-device (ED), and the co-channel SINR thresholds table to model the inter- and intra-SF interference~\cite{Magrin:ICC:2017}.
The simulation model includes the duty cycle limitations for frequency channels, channel hopping, and uplink and downlink transmission functionalities.
Three packet collision models are considered: (i) \textbf{B(P)}, baseline (pessimistic), where concurrent transmissions cause discarding of all packets.
(ii) \textbf{IC}, intra-SF collisions with capture effect, solely the concurrent transmissions with the same SF may cause packet failures, but the transmissions with different SF do not affect each other. The capture effect allows the gateway (GW) to receive the packet with the highest SINR for each SF and channel. (iii) \textbf{IIC}, transmissions with the same as well as with the other SF may cause packet discards. The capture effect between the same SF is enabled.

Two cases were simulated in \cite{Markkula:ICC:2019}. In the \textbf{N1} case, all EDs applied SF7, and B(P) and IC collision models were used. One hundred instances of two-hour-long simulations were run for each parameter set, and the results were averaged. In the \textbf{N2} case, the EDs operated with SF7-SF12 in a random manner (50 EDs per each SF), and all three collision models were investigated. One hundred instances of five-hour duration simulations were run for each set of parameters. The key parameters for the simulated cases are presented in Table~\ref{l_param}. 300 EDs were located on the circular area using the random radius from 0 to 13 km allowing for close to 100\% PDR when no collisions occur. The EDs transmitted packets (8-bit application layer payload), using the Poisson distribution with particular mean packet generation interval, to the GW that was located in the center of the simulated area. Traffic varied from 0.1 to 1 erlang (E). Note, the value of 1 E refers to the full utilization of the network capacity within a single channel. The number of EDs was such that the duty cycle limitation of 1\% was never exceeded.

\begin{table}
\centering
\caption{Key parameters for two simulated cases}
\begin{tabular}{l|c}
\hline
\textbf{Parameter} & \textbf{Value}\tabularnewline\hline
Number of nodes & 300 (end-device), 1 (gateway)  \tabularnewline\hline
End-device traffic & 0.1, 0.2, ..., 1 E  \tabularnewline\hline
Spreading factor & 7 (N1 case), 7, 8, ..., 12 (N2 case)  \tabularnewline\hline
Duty cycle & 1\% \tabularnewline\hline
Channel band width & 125 kHz  \tabularnewline\hline
Base frequency & 868.1  \tabularnewline\hline
Transmission power & 14 dBm  \tabularnewline\hline
Tx antenna gain & 0 dBi   \tabularnewline\hline
Antenna height & 3 m (end-device), 24 m  (gateway) \tabularnewline\hline
MAC  & ALOHA, random frequency channel \tabularnewline\hline
Channel hopping  & \multirow{4}{*}{Disabled} \tabularnewline\cline{1-1}
Retransmissions  &  \tabularnewline\cline{1-1}
Downlink transmissions  &  \tabularnewline\cline{1-1}
ADR and power control  &  \tabularnewline\hline
Activation: & by personalization \tabularnewline\hline
\end{tabular}
\label{l_param}
\end{table}

\section{Numerical Results}\label{sec:model}

Now we use the models presented in Section~\ref{sec:models} to investigate the performance limitations of LoRaWAN.

\subsection{Coverage}

\begin{figure}[tb]
\centering
\includegraphics[width=.95\columnwidth]{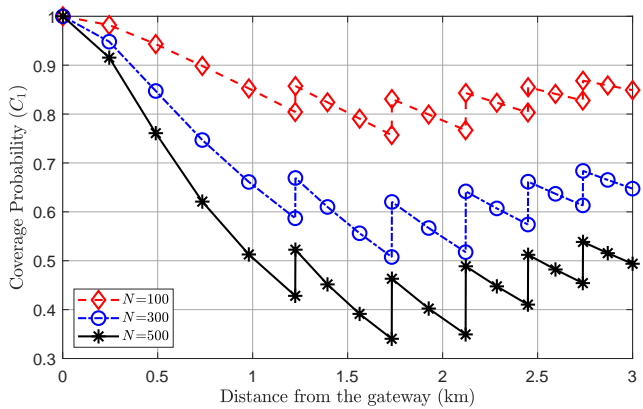}
\caption{Coverage probability of LoRaWAN.}
\label{fig:coverage}
\end{figure}

Figure~\ref{fig:coverage} demonstrates the coverage probability of LoRaWAN in the scenario presented in Figure~\ref{fig:nodes} for a single frequency channel.
We assume a typical EU configuration, \textit{i.e.}, the 868MHz ISM band, $B=125$kHz bandwidth, forward error correction rate of $4/5$, and transmit power $P_{tx}=14$dBm.
We also assume the path loss exponent $\eta=2.75$, interferers' duty cycle at 1\% ($p=0.01$), AWGN power $\sigma_w^2 = -174 + F + 10 \textup{log}_{10}(B)$, and $F=6$dB to be the receiver noise figure.

Figure~\ref{fig:coverage} shows results for the different numbers of interfering nodes.
One can see that the interference ($Q_1$) has a high impact on coverage probability. The increase in the number of nodes results in a decrease in the coverage probability.
In the figure, it is also visible that the increase of the SF rate increases coverage probability.
However, inside each SF, coverage probability decreases quickly with distance due to path loss.
As can be seen, the equal-area SF allocation method generates a stabler performance scenario in our model because it equalizes interference in all SF rings.
Note that it only happens here because we do not consider a specific application and assume that devices always use their entire $1\%$ duty cycle.
If network usage changes with SF, interference equalization will depend on the on-air packet time for each SF, as suggested by~\cite{Tiurlikova:ICUMT:2018}.
Since we consider uniformly distributed nodes, rings with the same area will have, on average, the same number of nodes.
It has been shown that equal-width SF allocation induces inferior performance for higher SF due to interference since the outer rings would have larger areas with more nodes~\cite{Hoeller:Access:2018}.
The path loss-based SF allocation method has shown stabler results~\cite{Hoeller:Access:2019}.

\subsection{Throughput}

\begin{figure}[tb]
\centering
\includegraphics[width=.95\columnwidth]{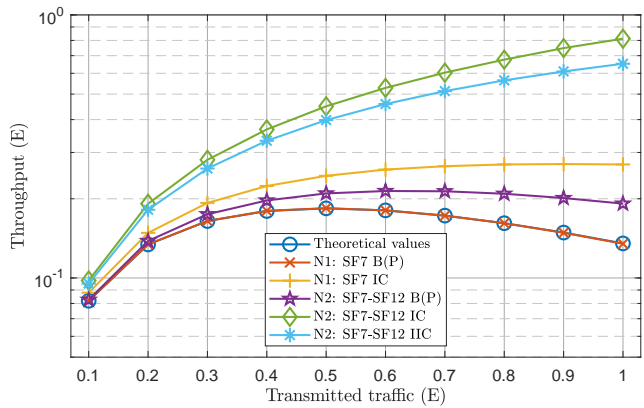}
\caption{Throughput as a function of transmitted traffic.}
\label{NET1}
\end{figure}

Figure~\ref{NET1} presents the results of the N1 and N2 simulation cases. The curves illustrate the throughput ($\textit{S}$) as a function of transmitted traffic ($\textit{G}$). In addition to the simulation results, the theoretical performance for pure-ALOHA ($\textit{S} = \textit{G}\textit{e}^{-2\textit{G}}$) is also plotted (the lowest curve). In N1, all the EDs operate with SF7 resulting in the on-air time for a single packet equal to 46.3ms. In N2, as in the analytical model discussed above, all the SF (7-12) are distributed equally between the devices. The average packet duration is approximately 399.5ms.

\begin{figure}[tb]
\centering
\includegraphics[width=.95\columnwidth]{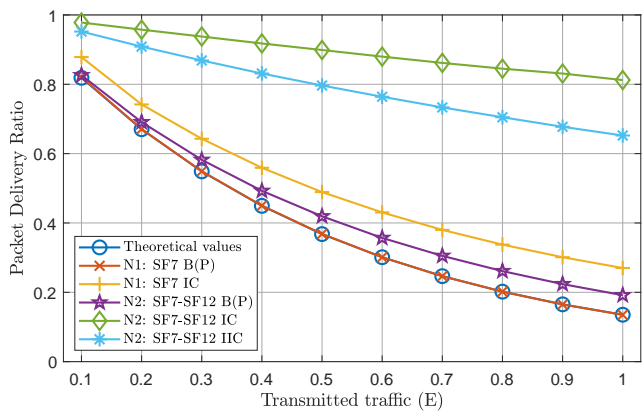}
\caption{Packet delivery ratio as a function of transmitted traffic.}
\label{NET2}
\end{figure}

The baseline (\textit{i.e.}, B(P)) results of the N1 case match closely with the theoretical results of pure-ALOHA (the difference is approximately 0.115\%). Accounting for intra-SF collisions and the capture effect (IC model) more than doubled the throughput, mainly due to the capture effect allowing for correct packet reception despite the interference.

By comparing the B(P) results of the N2 case with the theoretical values of pure-ALOHA, one sees that the throughput is higher in the N2 case. The difference increases, from 0.9 to 41.6\%, when the transmitted traffic amount is increased from 0.1 to 1~E. The phenomenon occurs because when increasing the traffic, the packets with shorter duration and lower SF values are more likely to be received without collisions than the lengthier transmissions with higher SF values.

Analyzing the results for the different collision modes, one sees that the IC model increases the maximum throughput by 18.4 to 323.7\% for the N2 case compared to B(P). Accounting for inter-SF interference (IIC model), the performance is further decreased by 3.1 to 83.7\% from the IC model (being from 15.3 to 240.3\% higher than the B(P) model).

The maximum network utilization with pure-ALOHA and N1 with B(P) model is 18.4\%, \textit{i.e.}, $\textit{S}$ is 0.184 E that corresponds to about four packets per second (p/s). The simulation results show that when multiple SFs were applied in N2, B(P) model resulted in maximum utilization of 21.4\% (about 1.6 p/s). The IC model with N1 featured the maximum utilization of 27\% (about 5.8 p/s), and N2 resulted in the utilization of 81.2\% (about 6.2 p/s). When also the inter-SF interference was considered (ICC model), the N2 case led to 65.2\% (about 5 p/s) maximum utilization of the network capacity. Thus, in the B(P) model, the highest number of packets could be delivered in the N1 case (2.4 p/s better than N2). In the IC model, N2 resulted in 0.4 p/s better performance, and in the ICC model, the N1 case enabled the delivery of 0.8 p/s more than the N2 case. Note that there is no difference between IC and ICC models for N1 because only a single SF is applied.

Figure~\ref{NET2} presents the PDR as a function of transmitted traffic. In all cases, PDR values decrease when the volume of transmitted traffic increases. The PDR values are from 82 to 13.5\% in the cases of pure-ALOHA and N1 with B(P). N2 with B(P) induced a bit higher PDR: from 82.6 to 19.2\%. Considering the capture effect and a single SF (IC model with N1) results in PDR values that are from 87.9 to 27\%. The curves illustrate that when multiple SFs were applied with the capture effect (ICC and IC with N2), the PDR decreased quite linearly when compared to the other results that had a steep decrease in PRD with the lowest transmitted traffic amounts. ICC and IC with N2 resulted in PDR values from 95.2 to 65.2\% and from 97.8 to 81.2\%, respectively. In N1 case, there were transmitted approximately 8.6 times more packets due to the higher average packet duration in the N2 case.              

\section{Discussions and Outlook}\label{sec:conclusion}

Compared to conventional and perspective cellular IoT, the current LPWAN technologies offer at least two clear benefits.
First, they use less signaling, which has a positive impact on latency, energy consumption, and device complexity and cost when network traffic is low or moderate. However, as we have shown, this approach has a drawback: in heavy-loaded networks, without efficient signaling, interference becomes a critical limiting factor for scalability.
Second, LPWAN does not implement handover mechanisms. It positively impacts the scalability and reliability of the multi-gateway networks (to be considered in further works), but introduces extra load to the backbone network and servers, implying additional costs.

The simulation results illustrate the maximum throughput values when using three different packet collision models and one or multiple SFs. A single frequency channel was used in the simulations. The maximum throughput would grow linearly with the number of applied channels. For example, the maximum throughput in IC with N1 (single SF), 0.27 E (5.8 p/s), is reached with the transmitted traffic of 0.8 E. With five-channel hopping, the maximum throughput would be 1.35 E (29 p/s), and the PDR would be 33.8\% in both situations, resulting the discard of 66.2\% of the packets due to collisions.

In future IoT applications, the number of active devices is expected to increase drastically, and interference will become a significant limiting factor.
In such scenarios, the reduced signaling of LPWAN technologies like LoRaWAN and SigFox can become a bottleneck, and the investment in backbone infrastructure to support this huge number of devices will increase.
Research efforts have started addressing such bottlenecks proposing more efficient and lightweight access control~\cite{Leonardi:IOTJ:2019} and adapting current cellular technologies like NB-IoT to the unlicensed spectrum~\cite{Sun:ISWCS:2019}. Meanwhile, there exists also a trend to simplify the signaling for particular data transfers in cellular technologies operating in licensed bands (\textit{e.g.}, the almost ALOHA-like early data transmission (EDT) for NB-IoT). As a result, a post-5G converged LPWAN connectivity will likely feature operations in both licensed and unlicensed bands with time- and frequency-division combination of ALOHA-like and grant-based channel access.

\section*{Acknowledgements}
This work or its authors have been partially supported in Finland by Academy of Finland (Aka) 6Genesis Flagship (Grant 318927), EE-IoT (Grant 319008), FIREMAN (Grant 326301), and Aka Prof (Grant 307492); and in Brazil by CNPq and CAPES-UFSC PrInt ``Automation 4.0'' program. 

\bibliographystyle{IEEEtran}
\bibliography{references}

\end{document}